\documentclass[twocolumn]{aastex631}

\let\tablenum\relax

\usepackage{natbib,epsfig,siunitx,url,graphicx,amsmath,footnote,float,xcolor,cases}
\usepackage{subcaption}
\usepackage{soul, xcolor}
\begin{document}

\title{On the local formation of the TRAPPIST-1 exoplanets}

\author[0000-0001-8933-6878]{Matthew S. Clement}
\affiliation{Johns Hopkins APL, 11100 Johns Hopkins Rd, Laurel, 20723, MD, USA}  
\affiliation{Earth and Planets Laboratory, Carnegie Institution for Science, 5241 Broad Branch Road, NW, Washington, DC 20015, USA}
\affiliation{Consortium on Habitability and Atmospheres of M-dwarf Planets (CHAMPs), Laurel, MD, USA}

\author[0000-0001-5272-5888]{Elisa V. Quintana}
\affiliation{Consortium on Habitability and Atmospheres of M-dwarf Planets (CHAMPs), Laurel, MD, USA}
\affiliation{NASA Goddard Space Flight Center, Greenbelt, MD 20771, USA}

\author[0000-0002-7352-7941]{Kevin B. Stevenson}
\affiliation{Johns Hopkins APL, 11100 Johns Hopkins Rd, Laurel, 20723, MD, USA}  
\affiliation{Consortium on Habitability and Atmospheres of M-dwarf Planets (CHAMPs), Laurel, MD, USA}

\correspondingauthor{Matt Clement}
\email{matt.clement@jhuapl.edu}

\begin{abstract}

The discovery of seven $\sim$Earth-mass planets, orbiting the 0.09 $M_{\odot}$ M-Dwarf TRAPPIST-1 captivated the public and sparked a proliferation of investigations into the system's origins.  Among other properties, the resonant architecture of the planets has been interpreted to imply that orbital migration played a dominant role in the system's early formation.  If correct, this hypothesis could imply that all of the seven worlds formed far from the star, and might harbor enhanced inventories of volatile elements.  However, multiple factors also contradict this interpretation.  In particular, the planets' apparent rocky compositions and non-hierarchical mass distribution might evidence them having formed closer to their current orbital locations.  In this paper, we investigate the latter possibility with over 600 accretion simulations that model the effects of collisional fragmentation.  In addition to producing multiple TRAPPIST-like configurations, we experiment with a number of different models for tracking the evolution of the planets' volatile contents and bulk iron-to-silicate ratios.  We conclude that a trend in bulk iron contents is the more likely explanation for the observed radial trend of decreasing uncompressed densities in the real system.  Given the degree of radial mixing that occurs in our simulations, in most cases we find that all seven planets finish with similar volatile contents.  Another confounding quality of the TRAPPIST-1 system is the fact that the innermost planets are not in first-order resonances with one-another.  By applying a tidal migration model to our most promising accretion model results, we demonstrate cases where higher-order resonances are populated.

\end{abstract}

\section{Introduction}
\label{sect:intro}

The TRAPPIST-1 exoplanets \citep{gillon16,gillon17} are widely considered to be the most observationally accessible nearby ($\sim$39.1 ly) transiting habitable zone planets, and the best candidate worlds for atmospheric characterization with current instrumentation \citep[e.g.][]{lustig19,trubet20}.  Out of the $\lesssim$20 detected planets with $R<$ 2.0 $R_{\oplus}$ that spend the majority of their orbits inside their host star's conservative habitable zones \citep{hz,hill22}, TRAPPIST-1e, f and g constitute half of the sample of such worlds with masses and radii that are constrained to be truly Earth-like.  While substantial quantities of observing time have been invested in pursuing atmospheric detection in the system via transmission spectroscopy \citep{dewit16,dewit18,gillon17,luger17,wakeford19} and photometric secondary eclipse measurements \citep{greene23}, such campaigns have thus far produced null results \citep[it is also possible that the planets' atmospheres were lost long ago, e.g.][]{vanlooveren24}.  In the absence of these observations, and given the obvious degeneracies of plausible internal compositions the planets might possess at their respective measured densities \citep[$\sim$4.1-5.5 $g/cm^{3}$ for the outer 3 planets:][]{grimm18,lienhard20,agol21}, theoretical modeling efforts \citep[e.g.][]{ormel17,dong18,schoonenberg19,horiogihara20,mandt22} continue to provide the basis for further characterization of the system.


In this paper, we turn our attention to the early formation and dynamical evolution of the TRAPPIST-1 planets.  In particular, several important constraints have been leveraged in past accretion modeling of the system to argue in favor of one particular evolutionary pathway \citep[pebble formation in the outer disk and subsequent inward migration:][]{ogihara09,ormel17,coleman19} over another \citep[in-situ via planetesimal accretion:][]{raymond07_mdwarf,hansen15,hoshino22}.  These include the multi-resonant dynamical architecture of the system \citep{luger17_res}, the planets' inferred water mass fractions \citep[WMF:][]{dobos19}, their similar masses and low eccentricities.

A known issue with the inward migration scenario is the fact that the inner three planets do not inhabit first order mean motion resonances (MMR, e.g. 2:1, 3:2 or 4:3, etc.).  Rather, they are respectively lodged in 8:5 and 5:3 commensurabilities \citep[third and second order,][]{luger17_res}.  While resonant chains are an expected consequence of orbital migration during the gas disk phase \citep{masset01,lee02,kley12}, simulations resoundingly demonstrate that first-order resonances where conjunction occurs at the same longitude once per resonant cycle are overwhelmingly the most common outcomes \citep{terquem07,pierens08,izidoro17}.  Thus, while the inner two resonances in TRAPPIST's 8:5,5:3,3:2,3:2,4:3,3:2 chain are a possible outcome for specific eccentric damping timescales \citep{charalambous21} or a recession in the disk's inner edge over time \citep{pichierri24}, they remain difficult to reconcile within the paradigm of migration-driven formation.  It is also important to note that the possibility of no resonance existing between planets b and c cannot be ruled out given the current baseline of just a few years of transit timing variations (TTV) observations \citep{teyssandier22}.  Indeed, the coincidental proximity to a resonance does not guarantee a pair of bodies is actually resonant \citep[for instance, Jupiter and Saturn's orbital period ratio is 2.49, yet the planets are not currently resonant, nor is it likely that they ever were in the past:][]{clement20_mnras}.  

 \citet{huang21} proposed an alternative scenario where the presence of a gas-free cavity during the gas-disk phase is responsible for the inner three planets avoiding capture in the 3:2 MMR.  It is also possible that the planets formed in a chain of exclusively first-order resonances, and were subsequently dislodged from commensurability via  orbital instability \citep{izidoro17} or tidal migration \citep{batygin13a}.  However, an orbital instability in the system would most likely dislodge all planets from resonance \citep{ray21}.  Similarly, \citet{brasser22} used tidal migration models to convincingly argue that it is not possible for a primordial chain of first-order resonances (3:2,3:2,3:2,3:2,4:3,3:2) to be reshaped into the modern orbital configuration \citep[see][for further analysis of the effects of migration on spin states within the system]{millholland23}. 

Here, we return to the possibility that late stage giant impacts displaced any primordial resonances in the system.  In contrast to \citet{brasser22}, we do not presuppose a post-disk phase resonant configuration.  Instead, we look at the spectrum of resonances that are populated when the TRAPPIST-1 planets undergo tidal migration after having first grown via planetesimal accretion \citep[e.g.][]{wetherill91}.  \citet{coleman19} contrasted a planetesimal formation model of the system with one that predominantly relied on pebble accretion and concluded that both scenarios are potentially compatible with the system's dynamical architecture.  More recently, \citet{childs23} used a code that accounts for pebble accretion, planetesimal collisions and collisional fragmentation to track the formation and compositional evolution of the system.  They concluded that the prevalence of late giant impacts on all planets likely implies that, even if the planets migrated substantially during their formation, all seven are still most likely rocky and desiccated.  In this paper we build on these finding by focusing closely on the core and water mass fractions (CMF and WMF) of consecutive planets \citep[that seem to trend towards lower total iron contents with increasing semi-major axis in interior structure models:][]{chambers13,grimm18,agol21}.  Recently \citep{schneeberger24}, it was proposed that this radial density trend is a consequence of the planets having formed in a relatively dry disk that was rich in phyllosilicate-bearing pebbles.  Moreover, this inflation of core sizes for interior planets is of increasing interest to many fields of astronomy and planetary science, given the similar nature of Mercury's composition \citep{hauck13,clement21_merc2,clement21_merc3} and recent detections of so-called ``Super-Mercuries'' \citep{adibekyan21,rodriguez23}.  We also examine the distribution of material scattered onto stable orbits in the system's exo-asteroid belt during the formation process \citep{ray21}. While an inner debris disk has not been detected in the system, \citet{marino20} estimated an upper limit of such a disk's mass within 4 au to be similar to that of the modern asteroid belt.

The goal of our work is not to disprove or contend with migration-driven pebble accretion models of the system's formation \citep[e.g.][]{ormel17,coleman19} models.  Instead, we intend to perform an initial investigation to broadly test whether the combined processes of late stage giant impacts and subsequent tidal migration can produce the 8:5,5:3,3:2,3:2,4:3,3:2 chain.  Through this process we also aim to provide the community with a database of possible accretion histories of a high-priority system for observation with next-generation telescopes like JWST that can be used as inputs for future geophysical and geochemical modeling of the worlds (e.g. atmospheric evolution or internal structure models).  As is an unfortunate necessity when modeling the formation of exoplanets, certain aspects of our simulations admittedly rely on crude assumptions and potentially unrealistic input parameters.  Throughout our paper, we discuss and examine how these caveats compare to those of other accretion models in the literature.

\section{Methods}

\subsection{Planet Formation Simulations}
\label{sect:meth_form}

The numerical accretion models presented in the subsequent sections utilize a modified version of the $Mercury6$ hybrid integrator described in \citet{chambers13}.  The code uses relations derived in \citet{lands12} and \citet{sandl12} to check if impact geometries fall in the fragmenting or hit-and-run regimes and, if so, distribute the left-over mass between a number of equal-mass fragments that are each larger than the user-defined minimum fragment mass \citep[MFM $=$ 0.005 $M_{\oplus}$ in this paper, see][for a justification of this particular setting]{clement18_frag}.  Newly generated fragments are ejected from the massive remnant object with $v\simeq$ 1.05 $v_{esc}$ at uniformly spaced vectors within the collisional plane.

The incorporation of collisional fragmentation into our models serves two purposes.  First, the high orbital velocities of short-period objects in the TRAPPIST-1 system implies that a larger fraction of collisions fall within the fragmentation regime than those in comparable simulations of the Earth's growth.  Studies investigating the consequences of imperfect accretion during the formation of the solar system's terrestrial planets have found that, while the process does lengthen accretion timescales and damp the orbits of growing planets \citep{chambers13}, these effects only occur at the 10-20$\%$ level \citep{clement18_frag,haghighipour22} and do not significantly shift the final statistical distributions of system properties \citep{deienno19}.  As exoplanet studies including fragmentation have predominantly focused on solar system analogs \citep[e.g.][]{quintana16} or long-term dynamical evolution \citep{esteves22}, the role of fragmentation in the local accretion of rocky worlds around low-mass stars remains largely unexplored.  Second, utilizing such a scheme allows us track the evolution of each body's core mass fraction (CMF) by assuming that each initial particle in the simulation is differentiated with an Earth-like core sizes (CMF$=$ 0.3).  When fragmentation events occur, our approach is to first generate fragments from the projectile's mantle material, followed by its core, the target's mantle and finally the target's core.

Our simulations are envisioned to commence when TRAPPIST-1's natal gas disk had mostly dissipated, and the majority of planet growth and migration had already taken place.  Thus, each of our models embed 30 large protoplanetary embryos \citep{wetherill93,koko_ida_96,koko_ida_98} within a disk of 1,000 smaller planetesimals \citep{youdin05,johansen15,draz18}.  The ratio of embryo total mass to total planetesimal mass is fixed at 1.0 in our models.  While this choice of mass distribution is largely motivated by solar system studies \citep[e.g.][]{chambers98,raymond09a} and likely not representative of the authentic initial conditions in TRAPPIST's primordial disk, past works varying these ratios \citep[e.g.][]{jacobson14,walsh19} found that they only have a minor affect on the ultimate system architectures.  For simplicity, in all of our models we set the inner and outer disk boundaries (0.01-0.1 au) to approximately encompass the modern observed radial distribution of planets in the system.  While the outer disk truncation radius is a fairly arbitrary assumption designed to maximize the probability of forming planets with semi-majr axes close to those of the real ones.  The inner truncation radius is losely consistent with the star's magnetic truncation radius \citep[e.g.][note that the stellar radius of TRAPPIST-1 is approximately 0.12 $R_{\odot}$]{frank92,ormel17}: 

\begin{equation}
	a_{in} = 0.01 \bigg( \frac{B_{*}}{180 G} \bigg)^{4/7} \bigg( \frac{R_{*}}{0.5 R_{\odot}} \bigg)^{12/7} \bigg( \frac{M_{*}}{0.1 M_{\odot}} \bigg)^{-2/7}
\end{equation}

Assuming a magnetic field strength of 600 G at the equator \citep{reiners10}, this equation yields $a_{in}=$ 0.002 au.  Semi-major axes are assigned to all bodies such that the solid surface density profile of the disk follows a profile of the form $\Sigma \propto r^{-\alpha}$.  As the structure of TRAPPIST-1's primordial solid disk is entirely unconstrained \citep{chiang13}, in this paper we test $\alpha=$ 1.0, 1.5, 2.0 and 2.5 with the goal of determining which values support the formation of a non-hierarchical or bi-modal planetary mass distribution.  Unlike the partitioning of mass between embryos and planetesimals, the choice of $\alpha$ can strongly affect the number of final planets and their mass distribution.  Of note, most studies of terrestrial planet formation in the solar system set $\alpha=$ 1.5 \citep[e.g.][]{raymond06,birnstiel12,clement18}.  However, certain models have argued that values of $\alpha$ as high as 5.5 might be responsible for the Earth-Mars mass ratio \citep{izidoro15,izidoro22_nat}.  Additionally, we account for the possibility that $\sim$planet-mass objects are lost via collision with the Sun in half of our runs by testing an initial total disk mass ($M_{Disk}$) of 8.0 $M_{\oplus}$ \citep[note that the total mass of the observed planets is 6.46 $M_{\oplus}$][]{lienhard20}.  To maximize the chances of forming 7 planets with similar masses to the real ones, the other half of our computations use $M_{Disk}=$ 6.5.  This is also consistent with the results of studies attempting to estimate the minimum mass extra-solar nebula for M-Dwarfs that argue massive disks with $\gtrsim$5.0 $M_{\oplus}$ in solids interior to 0.5 au might exist during the epoch of planet formation in these environments \citep[e.g.][]{gaidos17,carmens21_mmsn}.

We initialized 1,000 simulations by assigning eccentricities and inclinations to our embryos and planetesimals by sampling Rayleigh distributions ($\sigma_{e}=$ 0.02, $\sigma_{i}=$ 0.2$^{\circ}$), and drawing the remaining angular orbital elements from uniform distributions of angles.  Of these 1,000 runs, 369 became intractable.  Specifically, they produced numerous fragments in repeated sequences of collisions and would have taken an infeasible amount of time to complete.  The remaining 631 systems were integrated for 20 Myr, during which they formed between 3-9 planets with M$>$ 0.05 $M_{\oplus}$ (around the mass of Mercury).  We experimented with different minimum planet mass definitions and found that system multiplicity is not particularly sensitive to minimum masses less than 0.3 $M_{\oplus}$.  In fact, only 25 total planets (0.7$\%$ of all planets) in our sample have 0.05 $<$M$<$ 0.15 $M_{\oplus}$ (half the mass of the smallest TRAPPIST-1 planet).  We also note that our selection of integration time is consistent with the values used in other recent studies of planetesimal accretion at short orbital periods in the literature \citep[e.g.][]{raymond07_mdwarf,coleman19,clement22,sanchez22,sanchez24,childs23}, and futher justified by the fact that the median time of the last embryo-embryo collision in our sample is 1.55 Myr. In the subsequent sections, we compare the properties of these systems to those of the TRAPPIST-1 exoplanets.  In particular, we focus on 225 systems with six planets, 104 that form seven planets, and 24 that finish with 8 total planets.  When comparing the real system to our six-planet systems, we neglect the presence of TRAPPIST-1h.  Thus, we consider these runs marginal successes as they might represent a hypothetical evolutionary track where the last planets' orbit destabilized, causing it to merge with one of the interior worlds.  Similarly, we consider 8 planet systems marginal success if the inner or outermost planet is less massive than any of the real planets. 

\begin{table*}[]
    \centering
    \begin{tabular}{|c|c|c|c|c|c|c|}
    \hline
    Planet & a (au) & Mass ($M_{\oplus}$) & Density (g/cm$^{3}$) & Surf. Grav ($g_{\oplus}$) & CMF & WMF \\
    \hline
         TRAPPIST-1b & 0.0115 & 1.37 & 5.43 & 1.08 & 0.252 & $<$10$^{-5}$ \\
         TRAPPIST-1c & 0.0158 & 1.31 & 5.45 & 1.07 & 0.266 & $<$10$^{-5}$  \\
         TRAPPIST-1d & 0.0223 & 0.39 & 4.35 & 0.61 & 0.197 & $<$10$^{-5}$  \\
         TRAPPIST-1e & 0.0293 & 0.69 & 4.89 & 0.80 & 0.246 & 0.003 \\
         TRAPPIST-1f & 0.0385 & 1.04 & 5.01 & 0.93 & 0.201 & 0.0  \\
         TRAPPIST-1g & 0.0468 & 1.33 & 5.04 & 1.02 & 0.161 & 0.0072  \\
         TRAPPIST-1h & 0.0619 & 0.33 & 4.15 & 0.56 & 0.165 & 0.006  \\
    \hline
    \end{tabular}
    \caption{Planet parameters for the TRAPPIST-1 system reported in \citet{agol21}.  Surface Gravity is reported with respect to Earth's. Core Mass Fraction (CMF) is inferred using a fully differentiated MgSiO$_{3}$ mantle and an Fe core.}
    \label{tab:trappist}
\end{table*}

\subsection{Constraints}
\label{sect:constraints}

Table \ref{tab:trappist} summarizes a number of observed and interpreted properties of the system reported by \citet{agol21}; the most recent work in a series of observational and modeling campaigns by a number of different groups \citep[e.g.][]{gillon16,gillon17,lienhard20,agol21} that have collectively greatly reduced various uncertainties.  In particular, the planets' masses are all known to within $\sim$0.01-0.05 $M_{\oplus}$ (less than around the mass of Mercury), and semi-major axes are constrained to extremely high precision.  While internal structure models are highly degenerate in the absence of any compositional measurements of the bodies' surfaces or atmospheres, given the range of initial assumptions tested in \citet{agol21}, the general radial trends in CMF and water mass fraction (WMF) represent important plausible system properties for formation models to attempt to match.  Thus, we do not impose strict constraints on these parameters.  Rather, we simply compare the radial trends for these parameters in simulations, to those inferred for the real system.

The dynamics of compact, resonant systems such as TRAPPIST-1 are strongly regulated by the orbital period ratios between neighboring planets that determine which MMRs they inhabit as well as the strength of their mutual perturbations.  Thus, rather than analyzing the rate at which our simulations generate planets with precisely correct semi-major axes, we predominantly focus on the distribution of neighboring planet orbital period ratios (between 1.3-1.7 in the real system).  In this manner, we compare the cumulative distribution of orbital neighboring planet mass and orbital period ratios in our various simulation sets to those of the real system.

\subsection{Caveats and Assumptions}

In general, our study utilizes initial conditions that are fairly similar to those employed in models of the formation of the solar system's terrestrial planets \citep[e.g.][]{chambers98,raymond06,clement18_frag}.  However, it is important to recognize several key ways in which the in-situ rocky-planet formation environment around low-mass stars likely differs from that of the early solar system.  While we neglect the effects of nebular gas in our simulations, depending on how the efficiency of planetesimal formation \citep{youdin05,johansen15,lichtenberg21} varies over the disk's lifetime, it potentially played a much more important role in the formation of the TRAPPIST-1 planets. However, the inferred ages of iron meteorite parent bodies \citep[that predate the emergence of the major chondrite groups,][]{trieloff03,kruijer17} seem to imply that planetesimal formation occurred rather early in the solar system.  Similarly, certain features in extremely young, observed protoplanetary disks \citep[such as dust:gas mass ratios and annular structures, e.g.][]{ansdell16,segura_cox20} have been interpreted to suggest that large solid bodies are present within the first few hundreds of kyr of the disk buildup phase.  However, it remains largely unclear how these processes scale to the low-mass regime of TRAPPIST-1 \citep[see][for a recent review]{pinilla22}.  In particular, the radial extent and dust:star mass ratios of disks around low-mass stars remains poorly constrained \citep{vandermarel18,vandermarel22,kurtovic21}; making it difficult to extrapolate models around Sun-like stars down to the M-Dwarf mass-scale.  On the one hand, disk models around low mass stars are best able to match mm-fluxes when radial drift of dust is neglected \citep{pinilla13}, perhaps indicating a more prolonged phase of planetesimal formation.  In contrast, disk models around low mass stars \citep[e.g.][]{schoonenberg18,schoonenberg19} typically conclude that planetesimal formation is highly efficient since the water snowline is so close to the star.  However, these results are highly dependent on a variety of unconstrained disk parameters.  While the primary goal of our present work is to investigate the effects of collisional fragmentation \citep[and to contrast these outcomes with the results of][that assumed perfect mergers]{clement22}, we plan to compare the results of our gas-free simulations with comparable models including nebular gas effects \citep[e.g.][]{morishma10} in a forthcoming companion paper.

\subsection{Tidal Evolution Simulations}

To investigate the effects of tidal migration after the conclusion of the planet formation process, we performed follow-on modeling of certain systems that most closely resembled the real TRAPPIST-1 architecture using the \textit{Posidonius} code \citep{blanco-cuaresma17,bolmont20,blanco--cuaresma21}.  The code includes additional forces and torques \citep[an extension of the tidal model within the \textit{Mercury-T} code:][]{bolmont15} necessary to account for tidal effects, rotational flattening and general relativity.  The N-body portion of the code is derived directly from the popular \textit{WHFAST} integrator incorporated in the \textit{REBOUND} package \citep{rebound} The package also includes algorithms necessary to model stellar or giant planet evolution \citep{baraffe98,leconte11,leconte13,galletandbolmont17}, nebular gas and the stellar wind, however we do not include these additional phenomena in our simulations.  We expect the planets' eccentricities to damp tidally after their formation, with a characteristic timescale \citep[e.g.][]{goldreich66,batygin13a} :
\begin{equation}
    \tau_{e} = \frac{P}{21 \pi} \bigg( \frac{m}{M_{*}} \bigg) \bigg( \frac{R}{a} \bigg)^{-5} \bigg( \frac{k_{2}}{Q} \bigg)^{-1}
    \label{eqn:edamp}
\end{equation}

And, to first order in e,

\begin{equation}
    \frac{da}{dt} = -2 e^{2} \frac{a}{\tau_{e}}
\end{equation}

Here, $P$ is the orbital period of a planet with mass $m$, $R$ is the planet's radius, $k_{2}$ is its Love number and Q is the tidal quality factor.  In a preliminary set of simulations we tested a range of values for planetary and stellar dissipation in a number of systems with planet orbits just outside of major second and third order MMRs.  More specifically, we started with the nominal values reported in the work of \citet{bolmont20} and varied the planetary fluid and potential Love numbers up and down by two orders of magnitude.  We also experimented with the effects of changing the time lag ($\Delta \tau$) by as much as one order of magnitude in either direction.  While we briefly discuss the results of these experiments later in our manuscript, for the purposes of consistency with the past literature the majority of our complete suite of $>$100, 100 Myr simulations employ the nominal values from \citet{bolmont20}.  Specifically, we use $k_{2,p}=$ 0.299, $k_{2,*}=$ 0.307, $\Delta \tau=$ 712 s, and fluid Love numbers for rotational flattening of 0.307 and 0.9532 for the star and planet, respectively.  As expected given equation \ref{eqn:edamp}, in all cases eccentricities for the inner four planets damped to near zero within 10 kyr.  In the case of the outer three planets (f, g and h), circularization timescales were between $\sim$500 kyr - 10 Myr, depending on the dissipation parameters selected.  In section \ref{sect:tides} we discuss cases where tidal migration post-formation produced resonances similar to those apparently present in the real system.

\section{Results}
\label{sect:results}

\begin{table*}[]
    \centering
    \begin{tabular}{|c|c|c|c|c|c|}
    \hline
    Parameter & Real& $\alpha=$ -1.0 & $\alpha=$ -1.5 & $\alpha=$ -2.5 & C22 No-Frag  \\
    \hline
    $N_{pln}$ & 7 & 6.3 & 6.3 & 6.6 & 7.0 \\
    $m_{b}$ & 1.37 & 0.38 & 0.54 & 1.11 & 0.35 \\
    $m_{c}$ & 1.31 & 0.55 & 0.76 & 1.48 & 0.62 \\
    $m_{d}$ & 0.39 & 0.98 & 1.10 & 1.49 & 0.56 \\
    $m_{e}$ & 0.69 & 1.36 & 1.41 & 1.11 & 0.69  \\
    $m_{f}$  & 1.04 & 1.59 & 1.35 & 0.83 & 0.65 \\
    $m_{g}$ & 1.33 & 1.31 & 1.16 & 0.62 & 0.60 \\
    $m_{h}$ & 0.33 & 0.39 & 0.59 & 0.34 & 0.36 \\
    $P_{c}/P_{b}$ & 1.60 & 2.03 & 2.11 & 1.72 & 2.88 \\
    $P_{d}/P_{c}$ & 1.67 & 2.43 & 2.17 & 1.92 & 2.80 \\
    $P_{e}/P_{d}$ & 1.51 & 2.08 & 2.07 & 2.01 & 2.45 \\
    $P_{f}/P_{e}$ & 1.51 & 2.01 & 1.95 & 1.87 & 2.44 \\
    $P_{g}/P_{f}$ & 2.02 & 1.87 & 1.94 & 1.82 & 2.27 \\
    $P_{h}/P_{g}$ & 1.52 & 1.84 & 1.86 & 1.83 & 2.28  \\
    $CMF_{b}$ & 0.252 & 0.30 & 0.30 & 0.36 & N/A \\
    $CMF_{c}$ & 0.266 & 0.41 & 0.45 & 0.43 & N/A \\
    $CMF_{d}$ & 0.197 & 0.40 & 0.40 & 0.42 & N/A \\
    $CMF_{e}$ & 0.246 & 0.47 & 0.44 & 0.43 & N/A   \\
    $CMF_{f}$ & 0.201 & 0.45 & 0.43 & 0.38 & N/A \\
    $CMF_{g}$ & 0.161 & 0.45 & 0.42 & 0.30 & N/A \\
    $CMF_{h}$ & 0.165  & 0.30 & 0.35 & 0.30 & N/A \\
    $WMF_{b}$ & $<$10$^{-5}$ & 0.0023 & 2.6 x 10$^{-4}$ & 1.2 x 10$^{-5}$ & 10$^{-5}$ \\
    $WMF_{c}$ & $<$10$^{-5}$ & 0.027 & 0.0099 & 3.4 x 10$^{-4}$ & 10$^{-5}$ \\
    $WMF_{d}$ & $<$10$^{-5}$ & 0.054 & 0.032 & 0.0010 & 0.001 \\
    $WMF_{e}$ & 0.003 & 0.055 & 0.054 & 0.0011 & 0.1 \\ 
    $WMF_{f}$ & 0.0 & 0.051 & 0.055 & 0.040 & 0.1 \\
    $WMF_{g}$ & 0.0072 & 0.050 & 0.055 & 0.054 & 0.1 \\
    $WMF_{h}$ & 0.006 & 0.070 & 0.055 & 0.059 & 0.1 \\

    \hline
    \end{tabular}
    \caption{Summary of planet properties in our successful analog systems (i.e.: 6, 7 and 8 planet systems, see section \ref{sect:meth_form} for an explanation of how specific planet analogs are defined when $N_{pln}=$ 6)).  The various columns are as follows: (1) the planet parameter, (2) the observed or reference value for the actual system (these data are essentially reproduced from table \ref{tab:trappist}), (3-5) the results for our new suite of simulations considering $\alpha$ values of -1.0, -1.5 and -2.5, respectively, and (6) results from similar formation models around 0.1 $M_{\odot}$ host stars from \citet{clement22} that did not include a fragmentation model and used $\alpha=$ 1.5}.  Each reported value is the median of value for all analogs of the particular TRAPPIST-1 planet formed in the respective simulation set.
    \label{tab:results}
\end{table*}

Table \ref{tab:results} provides a summary of various planetary properties for analog systems that formed between six and eight planets with $m>$ 0.05 $M_{\oplus}$.  An example time evolution for a system that formed a particularly good representation of the real system's mass distribution is plotted in figure $\ref{fig:time_lapse}$.  The masses of the planets in this system in order of increasing semi-major axis are: 1.67 $M_{\oplus}$ ($m_{b}=$ 1.37 $M_{\oplus}$ in the real system), 1.64 $M_{\oplus}$ ($m_{c}=$ 1.31 $M_{\oplus}$), 0.45 $M_{\oplus}$ ($m_{d}=$ 0.39 $M_{\oplus}$), 0.85 $M_{\oplus}$ ($m_{e}=$ 0.69 $M_{\oplus}$), 0.38 $M_{\oplus}$ ($m_{f}=$ 1.04 $M_{\oplus}$), 0.81 $M_{\oplus}$ ($m_{g}=$ 1.33 $M_{\oplus}$) and 0.20 $M_{\oplus}$ ($m_{h}=$ 0.33 $M_{\oplus}$).  Moreover, our follow-on \textit{Posidonius} tidal evolution simulation for this particular systems placed the inner five planets in a 2:1,5:3,8:5,3:2 resonant chain (as compared to 8:5,5:3,3:2,3:2 for the real three inner planets).  If the innermost planet could be lost to the central star in some way, this simulation could provide an example of a potentially viable evolutionary pathway for the formation of the system's unique resonant architecture.  However, it is not clear how the other resonances in the system would be affected during such an engulfment. We provide additional details on these models in section \ref{sect:tides}.

\subsection{Bulk system properties}

Several trends emerge immediately upon closer inspection of table \ref{tab:results}.  While the mass of TRAPPIST-1h is well reproduced throughout our simulation set, the most challenging planet to replicate in mass is TRAPPIST-1d.  Indeed, all of our simulation sets tend to produce mass distributions that are best described as peaked, with the radial location of the peak moving closer to the central star with decreases in $\alpha$ \citep[consistent with the results of other embryo accretion models in the literature that varied this parameter:][]{izidoro15}.  Thus, from a mass perspective, our $\alpha=$ -2.5 batch of simulations is modestly more successful as 6-8 planet systems are frequently composed of two $\gtrsim$ 1.0 $M_{\oplus}$ planets, slightly smaller versions of planets e-g ($\lesssim$ 1.0 $M_{\oplus}$) and a small TRAPPIST-1g.  However, the ``bi-modal'' nature of the system's real mass distribution proved to be extremely challenging to reproduce.

\begin{figure*}
    \centering
    \includegraphics[width=.8\textwidth]{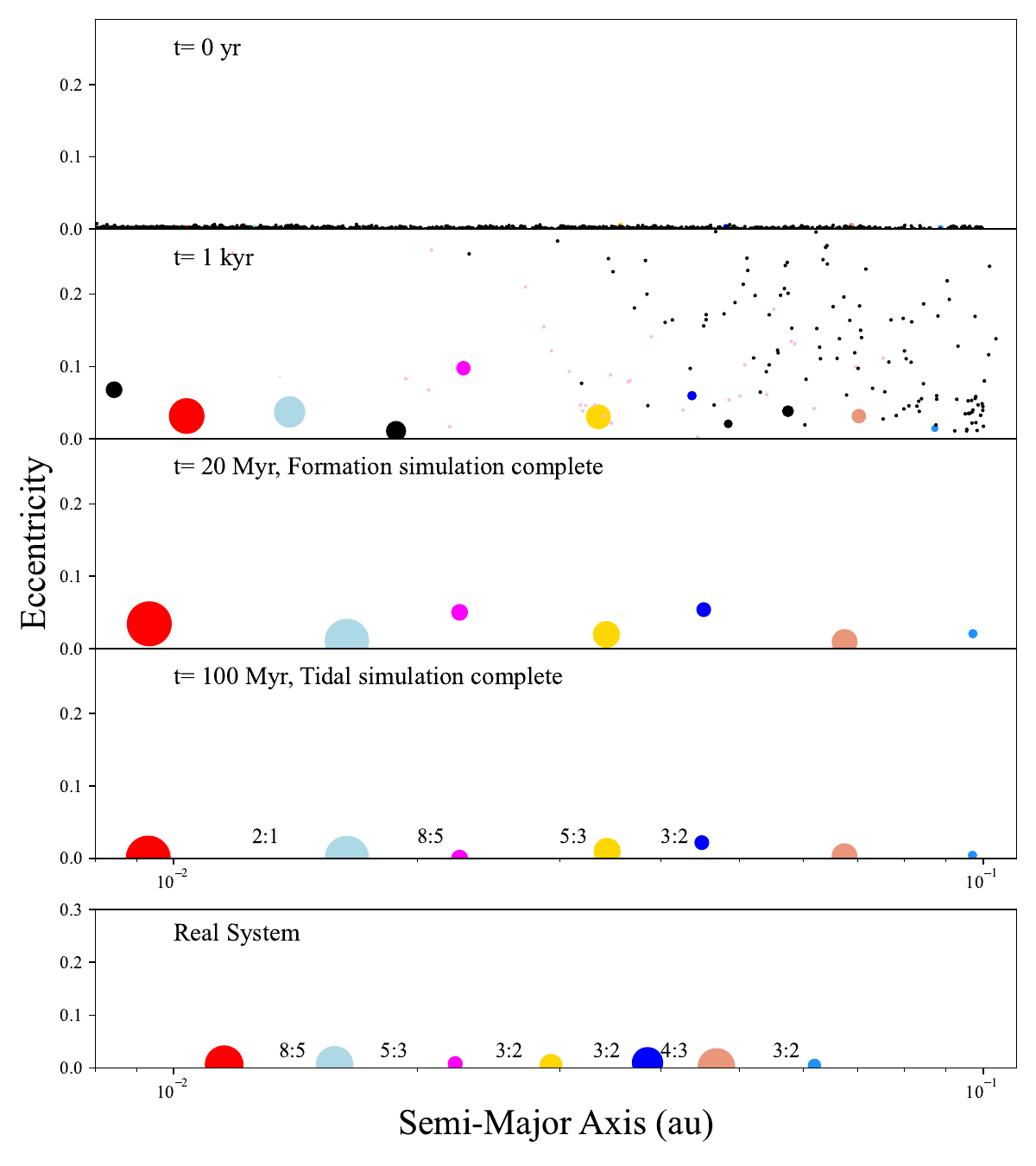}
    \caption{Example evolution of a system that was initialized with $\alpha=$ -2.5 and $M_{Disk}=$ 6.0 $M_{\oplus}$ that successfully formed a system of 7 total planets with orbits and masses very similar to those of the actual TRAPPIST-1 exoplanets (bottom panel). The semi-major axes and eccentricities of each body are plotted at each time interval, and the planets are color-coded to match their respective analog in the real system.  Additionally, it is worthwhile to note that our 100 Myr follow-on tidal evolution simulation (panel four) produced 4 total resonances (2:1, 8:5, 5:3, 3:2); labeled on the figure), two of which were higher order.}
    \label{fig:time_lapse}
\end{figure*}

While not the most typical outcome, it is possible to form a stable, smaller planet like TRAPPIST-1d in between neighboring sets of larger, $\sim$Earth-mass worlds.  To demonstrate this, we analyzed the frequency of forming planets at different positions in the system with $m<$ 0.5.  Among our seven and eight planet analog systems, 1.5\% contained small planets at position two (TRAPPIST-1c) with $\alpha=$ -2.5, as compared with 45$\%$ for $\alpha=$ -1.0. 
 The latter value is an obviously artifact of the fact that disks with shallower surface density profiles struggle to form large, $\sim$Earth-mass planets close to the central star.  Similarly, $\sim$1\% and $\sim$3\% of TRAPPIST-1d analogs were low in mass in our $\alpha=$ -2.5 and -1.0 simulation sets, respectively.  Figure \ref{fig:time_lapse} shows an example of such a system that simultaneously replicates the low masses of both TRAPPIST-1d and h. In contrast, small TRAPPIST-1e analogs were uncommon across all of our simulation sets (forming only in 1.5\% of our $\alpha=$ -2.5 simulations, and none of our other sets).

\begin{figure}
    \centering
    \includegraphics[width=.5\textwidth]{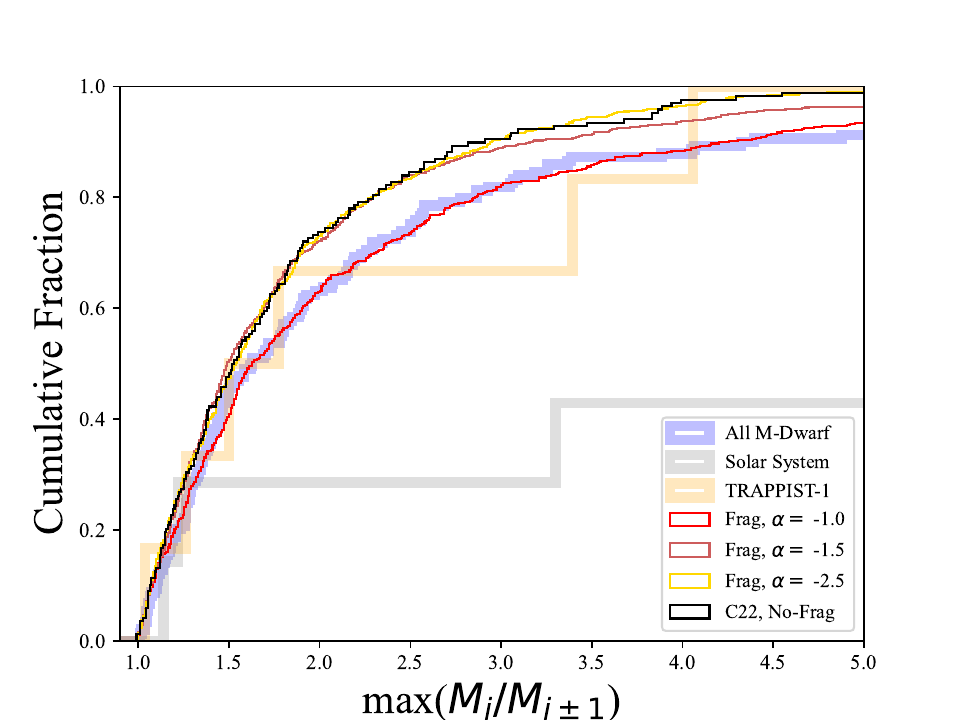}
    \caption{\textbf{The TRAPPIST-1 mass distribution is consistent with in-situ formation.} This figure plots the cumulative distribution of mass ratios of neighboring planets in the solar system (thick grey line), the real TRAPPIST-1 system (thick orange line), all known M-Dwarf-hosted multi-planet systems (thick blue line; compiled 1 May 2024), our simulations testing various values of $\alpha$ (red, black and gold lines for $\alpha=$ -1.0, -1.5 and -2.5, respectively), and similar simulations around a 0.1 $M_{\odot}$ star from \citet{clement22} that did not include a collisional fragmentation scheme (black line).  These lines only plot systems that form between 5 and 9 planets.}
    \label{fig:mrat}
\end{figure}

 While the system plotted in figure \ref{fig:time_lapse} (along with many others within our suite of runs) is an appealing analog of the real system's non-hierarchical mass distribution, we are most interested in measuring the rate at which our simulated planet systems replicate the fairly extreme mass ratios between neighboring pairs of TRAPPIST-1 planets.  Indeed, the large Mercury-Venus \citep[$\sim$ 14.8, e.g.][]{lykawka17} and Earth-Mars \citep[$\sim$9.3, e.g.][]{woo24} mass ratios in the solar system remain topics of intensive investigation, and seem anomalous when compared to most exoplanet systems that tend to harbor chains of planets with similar masses and radii \citep{weiss18,millholland21}.  In the TRAPPIST-1 system, mass ratios between neighboring planets range from near unity (b:c and f:g) to as high as 3-4 (c:d and g:h).  Figure \ref{fig:mrat} plots the distribution of neighboring planet mass ratios in our various simulation sets, compared with those of the real TRAPPIST-1 planets (thick orange line), all known M-Dwarf hosted multi-planet systems (thick blue line), and the solar system's planets (thick grey line, note that over half of the solar system's values are too extreme to fall between the plotted axes).  All data in the plot is reported as the ratio of the larger to smaller planet's mass, and thus should not be interpreted as evidencing any kind of radial trend in planet mass.

 Figure \ref{fig:mrat} demonstrates how the mass distributions produced in our various simulation sets are quite similar to those of the actual TRAPPIST-1 system, and other systems with multiple planets orbiting M-Dwarf hosts (defined here as $M_{*}<$ 0.6 $M_{\odot}$.  As is the case with many planetary system properties, when viewed in this manner the solar system is the clear outlier \citep[see][for a recent review on the orbital architecture of the solar system]{raymond24}.  However, it is important to remember that this figure is heavily biased by the absence of undetected planets with low masses or large orbital periods (and potentially large masses).  Nevertheless, it is interesting that our $\alpha=$ -1.0 simulation set provides a near exact match to both the thick blue (all M-dwarfs) and orange (TRAPPIST-1) lines.  The reason for this is that these systems are extremely likely to form an outermost planet that is much less massive than the second outermost planet.  When the surface density profile is steeper ($\alpha<$ -1.0), would-be diminutive outer planets dynamically couple more strongly to their nearest neighbor and are thus more likely to be lost via collision.  While this mechanism is advantageous in terms of consistently replicating the TRAPPIST-g:h mass ratio, the trend of large mass ratios existing almost exclusively between the most distant two planets does not extend to the entire exoplanet catalog.  Thus, the coincidental match between our $\alpha=$ -1.0 simulations' mass ratios and those of observed exoplanets does not lead us to strongly favor it over our other disks in our current analyses of the TRAPPIST-1 system. 

\begin{figure}
    \centering
    \includegraphics[width=.5\textwidth]{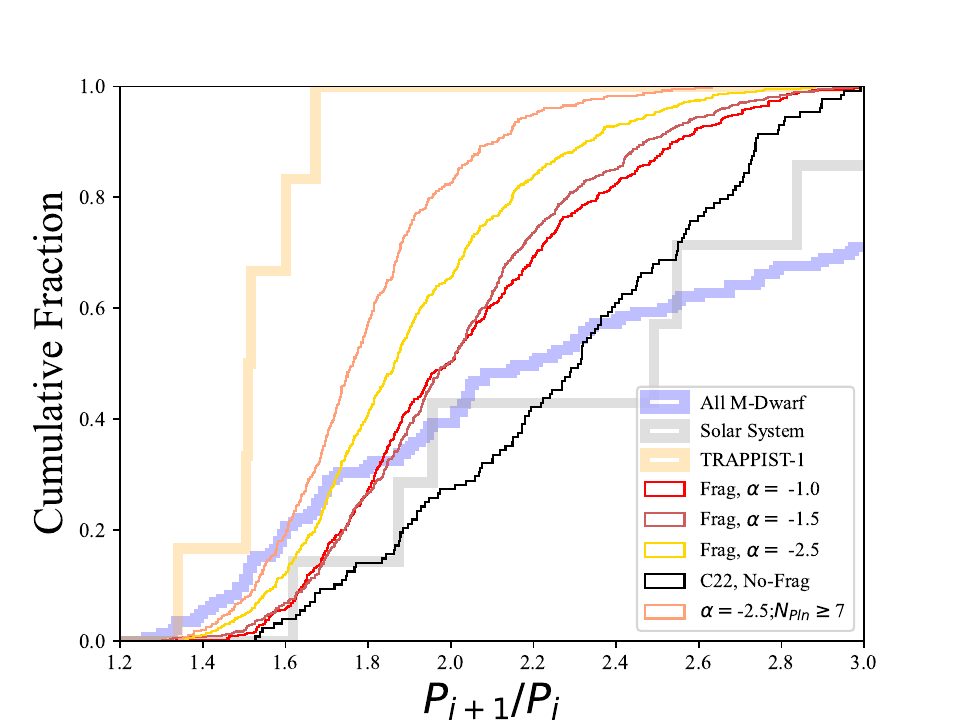}
    \caption{\textbf{The compact nature of TRAPPIST-1 is most consistent within a disk with a steep surface density profile.}  This figure plots the cumulative distribution of orbital period ratios of neighboring planets in the solar system (thick grey line), the real TRAPPIST-1 system (thick orange line), all known M-Dwarf-hosted multi-planet systems (thick blue line; compiled 1 May 2024), our simulations testing various values of $\alpha$ (red, black and gold lines for $\alpha=$ -1.0, -1.5 and -2.5, respectively), and similar simulations around a 0.1 $M_{\odot}$ star from \citet{clement22} that did not include a collisional fragmentation scheme (black line).  These lines only plot systems that form between 5 and 9 planets.  The salmon line, however, only depicts results from simulations with $\alpha=$-2.5 that formed seven or more planets.}
    \label{fig:prat}
\end{figure}

While the ratio of neighboring TRAPPIST-1 planet masses are well produced in our models, the distribution of orbital period ratios are not.  Replicating the distribution of planetary orbital period ratios for observed exoplanets has been a topic of considerable investigation in the recent literature \citep[e.g.][]{ogihara18,matsumoto20,izidoro21,esteves22}.  In one of the more popular models \citep{izidoro17}, the cores of short-period super-Earths grow rapidly in the outer disk via pebble accretion before they are transported inward into chains of MMRs via Type-I orbital migration \citep[e.g.][]{kley00,papaloizou06}.  This process produces a very steep distribution of orbital period ratios not particularly dissimilar from that of the TRAPPIST-1 system (figure \ref{fig:prat}, thick orange line).  Since these resonant chains are not intrinsically stable, a sizeable fraction experienced dynamical instabilities after the end of the nebular gas phase.  This second process flattens out the orbital period distribution into one that reasonably resembles the observed one (similar to the thick blue line in figure \ref{fig:prat}).  However, it remains unclear how well this process scales down to the regime of terrestrial planets orbiting M-Dwarfs.

As seen in figure \ref{fig:prat}, our simulation suites' distributions of final orbital period ratios fall somewhere in between those of the real system (thick orange line) and the full spectrum of M-Dwarf-hosted multi-planet systems (thick blue line).  This is unsurprising since we confine our initial embryos and planetesimals in a fairly narrow annulus that is roughly bounded by the modern orbits of TRAPPIST-1b and h, and do not allow large embryos to migrate via Type-I orbital migration.  Thus dynamical friction and scattering events between nearby embryos are the primary processes governing the semi-major axis evolution of our modeled proto-planets.  Through these processes, our initially confined disks tend to spread out \citep[e.g.][]{hansen09}.  However, it is worth noting that similar simulations including the effects of gas-driven migration also struggle to replicate the extremely compact nature of the TRAPPIST-1 system \citep[see, for example, figure 8 of][]{coleman19}.  As our $\alpha=$ -2.5 simulations were most successful at generating extremely compact systems of $\sim$Earth-mass planets (the salmon line in figure \ref{fig:prat} plots all $\alpha=$ -2.5 systems that formed seven or more planets), we conclude that ever steeper surface density profiles might be required to explain TRAPPIST's orbital architecture.

\subsection{Core Mass Fractions}
\label{sect:cmf}

The internal structures and compositions of planets provide inferences into the conditions of their formation \citep[e.g.][]{rubie15}.  It is challenging to constrain these properties in systems of known exoplanets, however future studies leveraging chemical models in tandem with transmission spectroscopy and accurate bulk density measurements have the potential to transform our understanding of planet formation in this manner.  In the case of TRAPPIST-1, our knowledge of these properties remains incomplete.  Nevertheless, the densities and corresponding surface gravities of each planet that are known with a fair degree of certainty are quite interesting for two reasons \citep{agol21}.  First, they are universally lower than those of the solar system's terrestrial planets.  This implies that they either have lower iron-to-silicate ratios, or higher volatile inventories than the Earth.  Second, the uncompressed densities of the planets decrease with increasing semi-major axis.  In the subsequent two sections, we use our library of accretion histories to investigate whether this quality is related to an underlying trend in iron or volatile contents (or both).

Recent studies of the formation of the solar system's terrestrial planets employing collisional fragmentation algorithms \citep{clement21_merc2,clement21_merc3,franco22,scora24,ferich24} have demonstrated that fragmentation events during particularly energetic giant impacts are a plausible explanation for the planet Mercury's large CMF \citep{benz88,benz07,asphaug14}.  However, it has also been suggested that various chemical processes might be responsible for the preferential formation of iron-rich planetesimals in the inner part of the terrestrial disk \citep[e.g.][]{ebelandalexander11,wurm2013,kruss20,johansen22}.  Comparative studies of exoplanet systems possessing high-density planets have the potential to break degeneracies between these two models.  Indeed, the apparent abundance of Super-Earth's with elevated CMFs \citep[occasionally referred to as ``Super-Mercuries,''][]{adibekyan21} strongly suggests that the processes responsible for altering Mercury's CMF likely operate in a variety of other systems as well.  While the masses  of the bodies involved in a collision that might alter the CMF of a super-Earth -- or one of our TRAPPIST-1 analog planets -- are similar to those of solar system simulations, the characteristic collisional velocities at short orbital periods are also necessarily much more extreme than those about Mercury's orbit.  In a recent study investigating the formation of super-Earths via giant impacts, \citet{dou24} concluded that a $\sim$5.0 $M_{\oplus}$ Super-Mercury could be formed in a 200 km/s collision between two 10.0 $M_{\oplus}$ bodies with a 45$^{\circ}$ impact angle.  While this velocity is obviously extreme, it is still somewhat comparable to the orbital velocities in the TRAPPIST-1 system (e.g. $\sim$80 km/s for TRAPPIST-1b).

We computed the final core mass fractions of each of our fully formed planets by post-processing their accretion histories using the methodology of \citet{chambers13}.  First, we initialized each embryo and planetesimal in the simulation with a CMF $=$ CMF$_{o}$.  We experimented with different values between 0.1 and 0.3.  In general, boosting the initial CMF affects the final CMFs of TRAPPIST-1f, g and h more strongly than those of the other planets because they tend to experience far fewer high-energy, CMF-altering collisions (thus they are the most likely to retain CMF $=$ CMF$_{o}$ through the entire simulation).  For these reasons, the subsequent text focuses on models where we set CMF$_{o}$ $=$ 0.20; the derived value of TRAPPIST-1f's CMF from \citet{agol21}.  When a fragmenting collision takes place, we assume that fragments are generated from the mantle of the projectile (CMF $=$ 0.0).  If the mantle is completely eroded, subsequent fragments are derived from the core of the projectile (CMF $=$ 1.0), followed by the mantle of the target, and finally the core of the target.  After a collision occurs, the remnant particles are assumed to differentiate instantaneously.  We experimented with different schemes (e.g. fragments being assigned even amounts of target and projectile material) and differentiation timescales and determined that the particular methodology does not strongly influence the distribution of final planetary CMFs in an ensemble of simulations.  Thus, we opted to present results with this fairly straightforward methodology in order to better facilitate comparisons with past studies \citep[e.g.][]{clement18_frag,clement21_merc3} that focused on the solar system.

\begin{figure*}
    \centering
    \includegraphics[width=.8\textwidth]{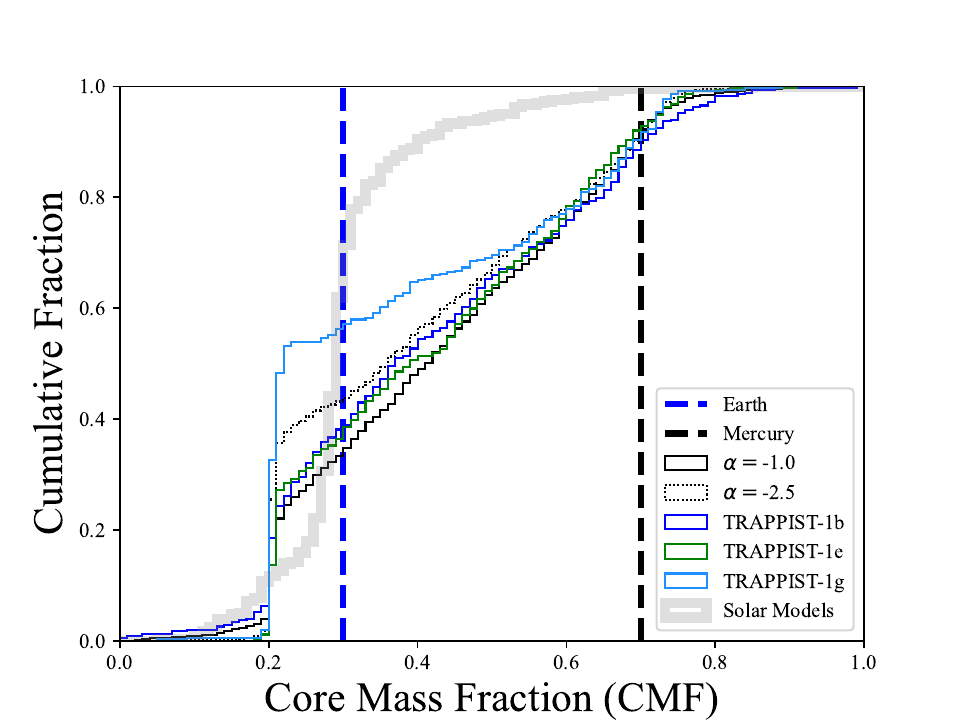}
    \caption{\textbf{High velocity collisions are more likely to alter the Fe-Si contents of the TRAPPIST planets than in similar simulations of accretion in the inner solar system.}  The figure plots the cumulative distribution of final CMFs for planets formed using different values of $\alpha$ (solid and dashed black lines), compared with all TRAPPIST-1b, e and h analogs (blue, green and light blue lines, respectively; see definitions in section \ref{sect:meth_form}).  Of note, TRAPPIST-1e has the highest bulk density and the third lowest mass of all the planets in the actual system \citep{grimm18}.  For reference, the thick grey line plots all Earth analogs (0.85 $<a<$ 1.20 au and 0.60 $<m<$ 1.5 $M_{\oplus}$) in solar system (SS) terrestrial planet formation simulations presented in \citet{clement18_frag} using the same collisional fragmentation scheme and settings.  The vertical blue and black lines represent the interpreted values for Earth and Mercury, respectively \citep{hauck13}.}
    \label{fig:cmf}
\end{figure*}

Figure \ref{fig:cmf} plots the distribution of fully accreted planet CMFs for our various simulation sets, and for particular planet analogs in our subset of simulations forming between 6 and 8 planets.  It is clear from the characteristic shapes of the curves corresponding to data from our new TRAPPIST-1 formation models that CMF-enhancement is far more common than diminution.  This trend is a consequence of the fact that collisional fragments -- which are most often composed of mantle material -- are extremely prone to loss via merger with the central star given the relatively large ejection velocities.  Thus, planets forming close to the central star are much more likely to bleed mantle material onto the star and end up with high CMFs than they are to acquire mantle-rich fragments that would drive their CMFs down.  This is also evidenced by the fact that the overabundance of planets with boosted CMFs compared to those with depleted cores is more pronounced for TRAPPIST-1b analogs (light blue line in figure \ref{fig:cmf}) than planet g (dark blue line analogs.  

To further illustrate the uniqueness of the collisional environment in the vicinity of the innermost TRAPPIST-1 worlds, we compared our computed CMFs to those derived in similar simulations of the solar system's formation \citep[thick grey line in figure \ref{fig:cmf},][]{clement18_frag}.  In these comparison models, $CMF_{o}$ is set to 0.3 in order to more frequently match the derived CMF of the Earth.  In these models, planets with highly-elevated CMFs are nearly as common as those with extremely reduced CMFs since collisional fragments are rarely ejected on to trajectories that are sufficiently extreme to lead to rapid merger with the Sun.

\begin{figure}
    \centering
    \includegraphics[width=.49\textwidth]{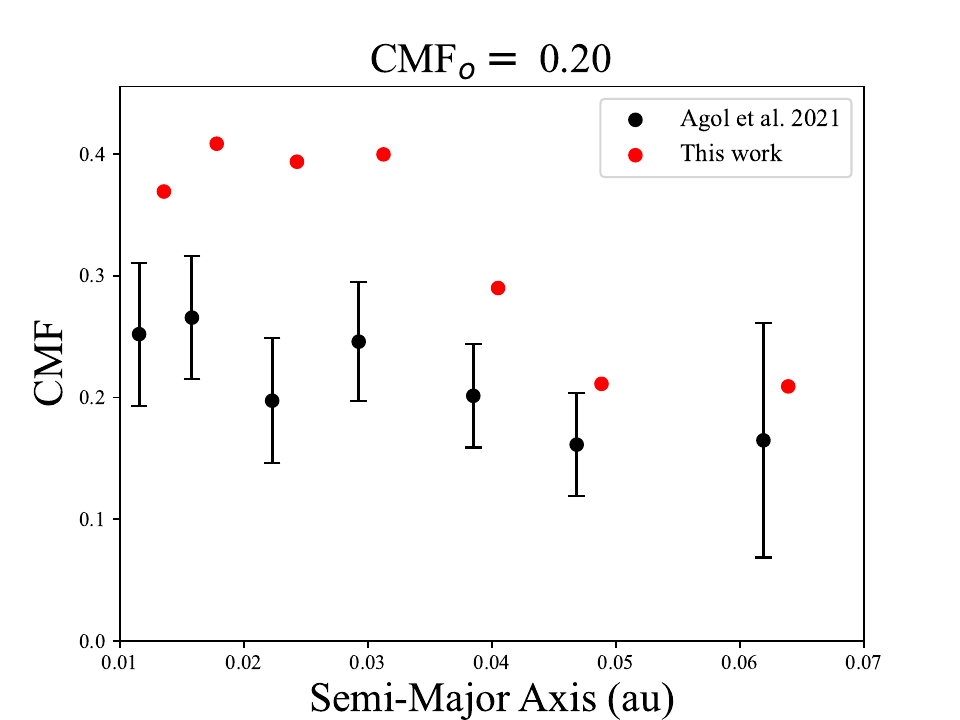}
    \caption{\textbf{A gradient in core mass fraction is a natural outcome of accretion in the TRAPPIST-1 system.}  This figure plots the comparison of median final CMF for TRAPPIST-1 planet analogs (red points, see definitions in section \ref{sect:meth_form}) formed in our current simulations that set $\alpha$ to -1.0, compared to the inferred values from the fully differentiated model of \citet{agol21} (black points and error bars).}
    \label{fig:cmf2}
\end{figure}

Generally speaking, each of our tested disk surface density profiles is capable of producing a hierarchical distribution of CMFs consistent with the observed trend in surface gravities in the real system (figure \ref{fig:cmf2}).  However, we were unable to develop a model capable of reproducing the relatively mild trend in CMF with increasing semi-major axis envisioned by \citet{agol21}.  This likely implies that the fragment ejection velocity ($\sim$1.05$v_{esc}$) in our simulations was too high, causing the efficiency of fragment re-accretion to be too low.  Thus, planets close to the central star bled mantle material onto the central star more efficiently than would be required to produce a good match to \citet{agol21}.  It is important to note here that our simulations did not include tidal damping or gas disk interactions.  While these dynamical damping mechanisms would not operate quick enough to save fragments ejected on star-crossing or nearly star-crossing orbits from loss, they could potentially prevent fragments ejected on less extreme trajectories from merging with the central star and ultimately reduce the final CMFs of the inner planets.  We also find that disks with shallower slopes tend to more efficiently enhance the CMFs of the innermost planets.  We suspect that this is a result of the lower relative density of planetesimals in the inner disk providing an insufficient degree of dynamical friction to damp the orbits of excited, mantle-only collisional fragments before they are lost via merger with the star.  However, because this trend is fairly weak, and the final CMFs of our inner planets are potentially too large (figure \ref{fig:cmf2}), we conclude that all values of $\alpha$ are viable from the perspective of CMF-alteration.

\subsection{Water Mass Fractions}

\begin{table*}[]
    \centering
    \begin{tabular}{|c|c|c|c|}
    \hline
    Model & WMF $a<$0.03 au (e/p) & WMF 0.03$<a<$0.08 au (e/p) & WMF $a>$ 0.08 au (e/p) \\
    \hline
        Trappist Reference & 10$^{-5}$/10$^{-5}$ & 10$^{-2}$/10$^{-2}$ & 0.1/0.1  \\
        Flat & 10$^{-5}$/10$^{-5}$ & 10$^{-5}$/10$^{-5}$ & 0.1/0.1 \\
        Desiccated All 1 & 0/0 & 0/0 & 0.1/0.1 \\
        Desiccated All 2 & 0/0 & 10$^{-5}$/10$^{-5}$ & 0.01/0.01 \\
        Icy & 10$^{-3}$/10$^{-3}$ & 10$^{-3}$/10$^{-3}$ & 0.25/0.25 \\
        Desiccated Planetesimal & 10$^{-3}$/0 & 10$^{-3}$/0 & 0.1/10$^{-3}$ \\
        Desiccated Embryo & 0/0.01 & 0/0.1 & 0/0.25 \\
        Evolve Embryo & 10$^{-3}$/10$^{-3}$ & 10$^{-3}$/10$^{-3}$ & 0.1/0.1 \\
        Evolve Planetesimal  & 10$^{-3}$/10$^{-3}$ & 10$^{-3}$/10$^{-3}$ & 0.1/0.1 \\
        \hline
    \end{tabular}
    \caption{Underlying initial volatile distributions for the various models depicted in figure \ref{fig:wmf}.  The columns are as follows: (1) the name of the model, (2) the initial WMF of embryos (e) and planetesimals (p) in region 1 (inside the modern habitable zone), (3) the initial WMF of bodies in region 2 (between the modern habitable zone and the water-ice snowline) and (4) the starting WMF in region 3 (outside of the snowline).}
    \label{tab:wmf}
\end{table*}

We followed the methodology of \citet{lichtenberg22} to compute potential distributions of planetary WMFs in our systems via post-processing of their accretion histories.   As with our CMF analyses in the previous section, our goal here is to treat the initial WMF gradient as a free parameter (within reason), and search input profiles that might explain the current system's distribution of uncompressed densities.  It should be noted that -- given the degeneracies involved in any particular underlying planetary bulk compositional model we might choose -- there is no real way to constrain this analysis on a planet by planet basis.  While we compare our results to the best fit water contents from \citet{agol21}, we are more interested in evaluating whether the degree of radial mixing in our models is small enough to allow a given initial radial distribution to sufficiently persist through the accretion process.  To keep our analyses roughly consistent with the current state of knowledge of the initial volatile lay down of solid bodies in protoplanetary disks as inferred from solar system meteorites and disk chemistry models \citep[e.g.][]{meech19}, we assume that our initial disks contain three separate reservoirs of material with distinct initial volatile contents:

\begin{enumerate}
    \item Inside of the modern habitable zone \citep[$\lesssim$ 0.03 au, e.g.][]{hz,hz2}
    \item Between the habitable zone's inner edge and the location of the water-ice snowline \citep[$\sim$ 0.08 au, e.g.][]{ida04}
    \item Outside of the water ice snowline.
\end{enumerate}

\begin{figure*}
    \centering
    \includegraphics[width=.95\textwidth]{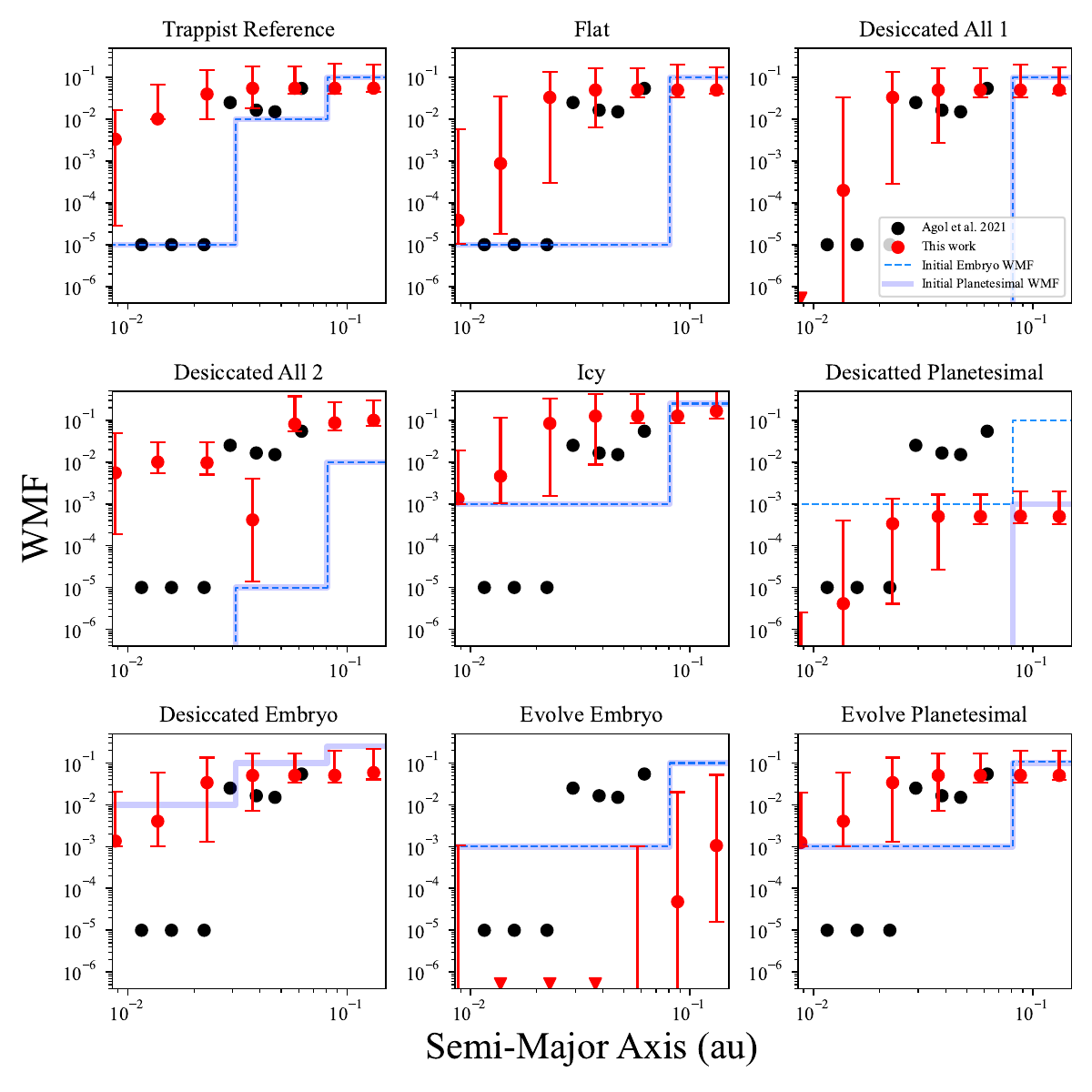}
    \caption{\textbf{None of the initial volatile gradients tested in this work can explain the observed gradient of uncompressed densities in TRAPPIST-1.}  Each panel provides the median final WMF for all TRAPPIST-1 planet analogs (red points and error bars, see definitions in section \ref{sect:meth_form}) formed in our current simulations, compared to the inferred values from  \citet{agol21} (black points, assuming CMFs of 0.25 for planets b, c and e, and 0.18 for the other planets) for a range of different initial volatile gradients (blue and light blue lines).  The error bars enclose 95$\%$ of the results from our complete simulation set.  While the black points for the real system are plotted at the location of the observed planet semi-major axes, the red points for our simulated planets are positioned at the average semi-major axis of all planets encompassed by the data point.}
    \label{fig:wmf}
\end{figure*}

In a series of experiments, we post-processed the accretion histories of all of our systems that formed between 6-8 total planets in order to estimate the final WMF of each fully accreted planet.  We tested over a hundred different possible initial volatile distributions and desiccation schedules.  In the subsequent text we expand on a subset of nine of the more interesting models.  The initial conditions for these experiments are summarized in table \ref{tab:wmf}, and the WMFs of each TRAPPIST-1 analog planet in each model is plotted in figure \ref{fig:wmf}, along with our nominal comparison case from \citet{agol21} that assumes each planet has CMF$=$0.25.

In collisions that are perfectly accretionary, the single remnant retains all the water from both the target and projectile.  However, our methodology for handling fragmentation when post-processing accretion histories diverges slightly from that of our CMF analyses in the previous section.  When a fragmenting collision occurs, we assign water from the colliding system to fragments in a manner that is proportionate to their mass times a presumed fragmentation efficiency, i.e.: 
\begin{equation}
    WMF_{Frag} = \eta_{Frag} \frac{WMF_{Targ}M_{Targ} + WMF_{Proj}M_{Proj}}{M_{Targ} + M_{Proj}}
\end{equation}
The remaining water is assigned to the collision's largest remnant.  We tested values of $\eta_{Frag}$ between 0-1.0.  In general, increasing $\eta_{Frag}$ tends to suppress the WMFs of the inner three planets since it boosts the rate at which volatiles are dumped onto the central star via fragments, and does not strongly affect the final WMFs of the outer planets.  The models presented in figure \ref{fig:wmf} depict scenarios with $\eta_{Frag}=$ 0.75.  This is based off the results of studies investigating volatile reaccretion after high-velocity giant impacts \citep[e.g.][]{gladman09} that show the tendency of the original target to quickly accrete disrupted volatiles.  We also experimented with the effects of temporally varying the volatile contents of objects.  In the two models labeled "Evolve" in figure \ref{fig:wmf} and table \ref{tab:wmf}, we assume either the embryo population (including the planets themselves) or the planetesimal population in region 1 lose all their volatiles (WMF$=$0.0) at $t=$ 100 kyr, and that bodies in region 2 desiccate at $t=$ 1 Myr.

The first volatile profile in figure \ref{fig:wmf} (Trappist Reference) essentially initializes each region of the disk with a WMF consistent with the inferred values of the constituent planets in the modern system \citep[as taken from our comparision case:][]{agol21}.  The fact that each final planet possesses a WMF in excess of the initial WMF in its region necessarily implies a reasonable amount of radial mixing in our simulations.  The degree of enhancement is large, but also isn't particularly surprising.  Indeed, the addition of just one embryo ($m\simeq$ 0.10-0.13 $M_{\oplus}$) from outside of the snowline (WMF$_{o}=$ 0.1) is enough to double the WMF of an Earth-mass planet with WMF$_{o}$= 0.01.  Similarly, fragmenting collisions occur with a reasonable frequency throughout the TRAPPIST disk, and thus provide an additional efficient means of boosting the WMFs of growing planets, even for $\eta_{Frag}=$ 0.75.  While this effect is less extreme in our steeper disks (see table \ref{tab:results}) since the inner disk is more dynamically isolated from the outer disk, it is still pronounced.  To isolate this effect, our next volatile profile (flat) assigns embryos and planetesimals in both region 1 and 2 $WMF=$ 10$^{-5}$, and assumes that objects outside of the snowline are endowed with a relatively high WMF of 0.25 \citep[similar to that of carbonaceous asteroids in the main belt like Ceres:][]{mcsween18}.  The large error bars for the inner three planets illustrate how the stochasticity of the process of water delivery from beyond the snow-line increases closer to the central star.  While this model provides one of our better matches to the \citet{agol21} WMFs, it is still unable to consistently prevent icy planetesimals and embryos from elevating the WMFs of the innermost planets.  This is potentially most problematic for reconciling the densities of the larger inner planets (b and c, $\simeq$ 5.5 g/cm$^{3}$) with those of the larger outer planets (f and g, $\simeq$ 5.0 g/cm$^{3}$).

We also experimented with a number of models that initialized disk regions with $WMF=$ 0.  Of these, ``Desiccated All 2'' provides perhaps the best match to our comparison case from \citet{agol21}.  However, the number of simulations where planets c and d attain very large WMFs by accreting material from outside of the snow line is still quite high.  The models ``Desiccated Planetesimal'' and ``Desiccated Embryo'' demonstrate how the final planets' WMF is, unsurprisingly, mostly set by the objects (either embryos or planetesimals) in the local region of the disk with the highest WMF.  Finally, the results of our two models that employ a desiccation schedule (``Evolve'') diverge from one another since desiccating embryos over time also serves to null the WMFs of mostly-accreted planets, while temporal desiccation of planetesimals only nulls the WMFs of a declining population of bodies.

In spite of all efforts, we were unable to produce a good match to the constant-CMF WMF distribution of \citet{agol21}.  Similarly, we were unable to produce a good match in section \ref{sect:cmf} by only manipulating the planets' CMFs. However, the problems we encountered in these two endeavours were different in kind.  In the case of estimating the distribution of final system WMFs, even for the most extreme and unrealistic initial conditions we tested, we were unable to produce a WMF gradient that was steep enough to explain the density dichotomy between the inner and outer planets.  However, it is possible that subsequent interactions with the central star could further reduce the planets' WMFs via photoevaporation \citep{lammer03} in a manner that might produce a better overall match to the \citet{agol21} values.  When it comes to CMF, we were able to consistently produce a CMF dichotomy, however it was always too extreme.  While the latter problem might simply imply that collisional fragmentation is not as efficient as considered here \citep[perhaps the typical ejection velocities are lower than assumed in our model, e.g.][]{emsenhuber24}, there is not an immediately obvious solution to the former problem.  However, it is also possible that a combination of both processes, or a mechanism not envisioned here, was responsible for sculpting the compositions of the TRAPPIST-1 planets.

In a similar study including the effects of collisional fragmentation, \citet{childs23} showed how the characteristically high impact energies of giant impacts occurring at such short orbital periods in the system might have vaporized would-be oceans \citep[e.g.][]{stewart14} on all seven planets (their figure 7).  We inspected the impact energies of all of our systems in a similar manner and found comparable distributions to those reported in \citet{childs23}.  However, our larger sample of systems with 6-8 total planets also contained many examples of systems that experienced relatively quiescent formation histories (about 1/3 of all evolutions) containing no impacts energetic enough to fully vaporize an Earth ocean of water.  In a companion study (Chen et al. in review) we use our suite of formation models in tandem with a planetary volatile evolution model that accounts for impact erosion, atmosphere-mantle exchange, outgassing, and a realistic planetesimal size distribution \citep{chen22} to investigate the potential final volatile distributions of our systems in much greater detail.  Additionally, it is important to point out that the very energetic collisions experienced by some of our system's could lead to the catastrophic destruction of one or more of the proto-planets.  It is unclear whether a second generation of planets could form in a system like TRAPPIST-1 after a high-energy collision grounds the growing planets down to dust.

\subsection{Asteroid Belt analogs}

\begin{figure}
    \centering
    \includegraphics[width=.49\textwidth]{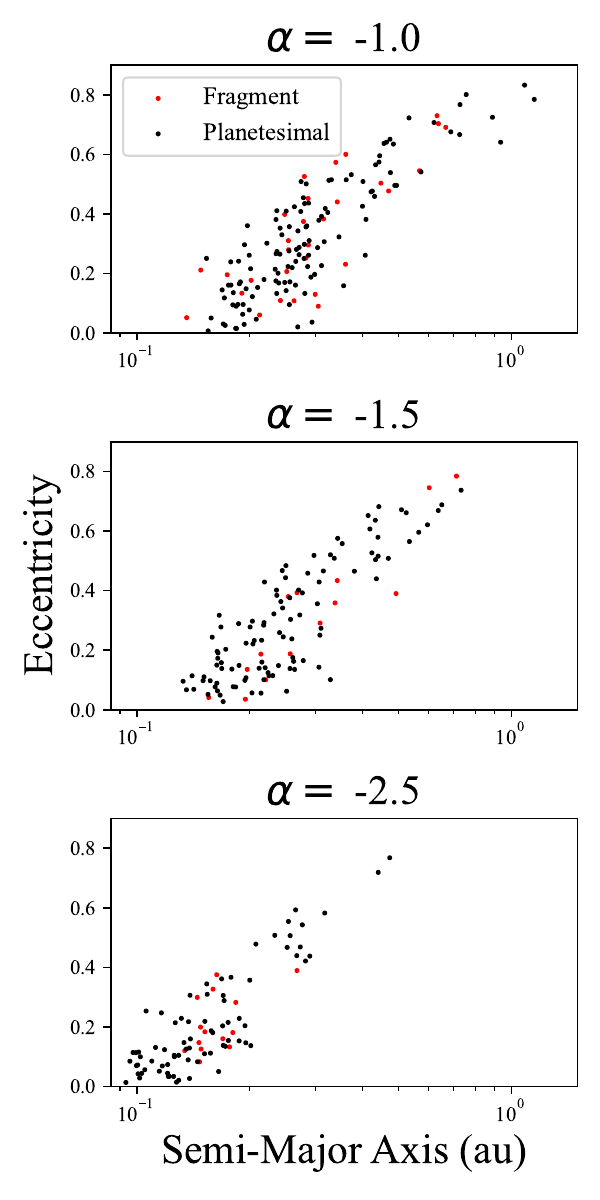}
    \caption{\textbf{Asteroid belt formation occurs as a consequence of the formation of the TRAPPIST-1 planets.} This figure provides the distribution of semi-major axes and eccentricities of surviving collisional fragments (red points) and planetesimals (black points) in the hot debris disk (exo-asteroid belt) region of our TRAPPIST-1 analog systems.  The three panels separate our systems formed from disks with different initial values of $\alpha$.  The figure only displays objects with semi-major axes greater than the most distant planet at the end of each simulation.}
    \label{fig:ab}
\end{figure}

It is widely accepted that some fraction of the objects in the solar system's modern asteroid belt formed in the vicinity of the inner planets before being transported into the belt through dynamical interactions that occur during the planet formation process \citep{bottke06_nat,ray17}.  This is supported by planet formation simulations \citep{izidoro16,izidoroetal24,deienno24} and chemical analyses of Enstatite meteorites that seem to have originated from asteroids in the inner belt \citep{avdellidouetal22,avdellidouetal24}.  Our TRAPPIST-1 simulations also finish with debris that was ejected from the planet formation region and stranded on stable, often high-$e$ and high-$i$ orbits beyond that of TRAPPIST-1h.  Figure \ref{fig:ab} plots the orbital distributions of these particles in our models testing different values of $\alpha$.  In general shallower disks produce larger inventories of surviving asteroids, while our $\alpha=$ -2.5 disks finish with fewer bodies as a result of their being initialized with less mass in more distant radial bins.  Specifically, the average total mass of leftover objects with orbital semi-major axes greater that that of the most distant planet in TRAPPIST-1 analogs finishing with between 6-8 planets in our models is 0.0035 $M_{\oplus}$ for our $\alpha=$ -1.0 disks, 0.0033 $M_{\oplus}$ for our $\alpha=$ -1.5 disks, and 0.0017 $M_{\oplus}$ for our $\alpha=$ -2.5 disks.  This implantation efficiency is relatively high when compared to models of terrestrial planet formation in the solar system by at least two orders of magnitude \citep[][indeed, the total mass of the modern asteroid belt is only around $\sim$ \num{5e-4} $M_{\oplus}$]{izidoroetal24}.  This discrepancy is a result of the larger total mass of our planet-forming disk relative to terrestrial planet formation models that are typically only initialized with $\sim$ 2.0 $M_{\oplus}$ in embryos and planetesimals \citep{izidoro22_nat}, and the much larger absolute surface density or solid bodies in our simulations (solar system models distribute bodies over a much broader range of semi-major axes).

The total masses of our TRAPPIST-1 asteroid belts are also well in excess of the observationally constrained maximum potential mass of a still undetected hot or cold debris disk in the system.  \citet{marino20} used ALMA observations to place an upper limit of \num{1e-5} $M_{\oplus}$ for a debris disk inside of 4 au (the vast majority of our surviving debris have a $<$ 1.0 au), and \num{1e-2} $M_{\oplus}$ for within 100 au.  There are several possible resolutions to this issue that we hope to explore in future work.  First, the inventory of objects in a TRAPPIST-1-hosted exo-asteroid belt would have declined through collisions and dynamical scattering events in the time since its formation.  The solar system's asteroid belt is thought to have lost around half its mass in the manner since the time of its formation \citep{minton11}.  Given that TRAPPIST-1 is much older than the Sun, it is certainly possible that an initially dense belt could be depleted by much more than a factor of two over the age of the system.  It is also possible that, through this process, a distant asteroid belt filled with icy bodies could continuously replenish the atmospheres of the system's planets via impacts \citep{kral18,clement22}.  Secondly, the presence of additional distant planets could also serve to limit and/or deplete a primordially large asteroid belt, both during and after its formation \citep{deienno22}.  The efficiency of these processes necessarily scale with the mass of the perturbing body.  However, the existence of the system's resonant chain \citep{raymond22} radial velocity data \citep{hirano20}, transit timing variations \citep{agol21} and astrometric constraints \citep{boss17} all place limits on the masses and potential orbital locations of any undetected planets.  Finally, it is also possible that we overestimate the fragment portion of the asteroid belt since we assume that all the material ejected after a fragmentation event is in the form of massive bodies, rather than dust \citep[e.g.][]{emsenhuber24}.  Thus, given the lack of evidence for the existence of an undetected eighth planet in the system, it seems reasonable to conclude that the majority of erosion of TRAPPIST-1's hypothetical asteroid belt would be due to mutual encounters between asteroids, and interactions with the known planets.  Finally, it is also possible that our formation models over-estimate that amount of mass deposited in the asteroid belt region.  This could be a result of our initial conditions (disk structure and mass partitioning) or our implementation of collisional fragmentation and hit-and-run collisions (in the sense that our model might be allowing too many small bodies to survive or be produced in collisions with growing proto-planets).  Additionally, if large scale orbital migration played a significant role in the formation of the system \citep{ormel17,coleman19}, it could have wiped out any primordial exo-asteroid belt \citep[e.g.][]{walsh11}.

\subsection{Subsequent tidal migration}
\label{sect:tides}

We performed a series of follow-on simulations investigating the tidal evolution of our 104 systems that possessed exactly 7 planets after the initial planet formation simulation using the \textit{Posidonius} code \citep{blanco-cuaresma17,bolmont20,blanco--cuaresma21}.  We experimented with a range of different initial values for the applied time lag, as well as both the stellar and planetary fluid and potential Love numbers.  The results presented in this section all utilize the reference values from \citet{bolmont20}.  In general, we noted that overly enhancing the tidal forces through excessive increases in the Love numbers tended to result in planet pairs migrating past both first and second order mean motion resonances with only short or no epochs of resonant evolution.  Even in our more extended suite of simulation using more moderate parameters, we observed many cases where two planets briefly become entrapped in a higher order resonance like the 8:5 or the 5:3 before quickly falling out of the commensurability.

\begin{figure}
    \centering
    \includegraphics[width=.49\textwidth]{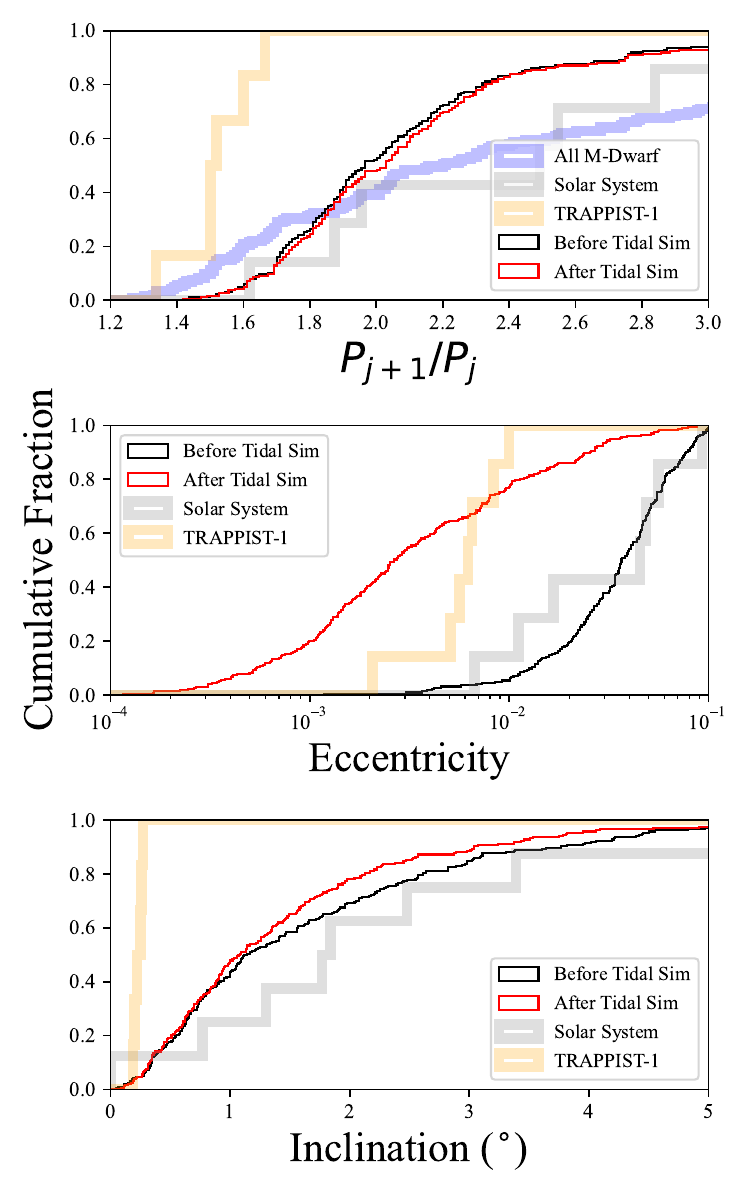}
    \caption{\textbf{Tidal evolution damps post-formation eccentricities.} The various panels plot the cumulative distribution of system orbital parameters (\textbf{Top}: orbital period ratios of neighboring planets, \textbf{Middle}: Eccentricity, and \textbf{Bottom}: Inclination) before (black lines) and after (red lines) our tidal evolution simulations (section \ref{sect:tides} compared with the solar system (thick grey line), the real TRAPPIST-1 system (thick orange line), all known M-Dwarf-hosted multi-planet systems (thick blue line; compiled 1 May 2024). }
    \label{fig:tides}
\end{figure}

Through these additional simulations we confirmed that, while the final eccentricities in our initial \textit{Mercury6} simulations are often much larger than those of the real planets \citep{grimm18,agol21}, the tidal circularization timescales for each planet are quite rapid.  The top panel of figure \ref{fig:tides} shows how the distribution of orbital period ratios does not change substantially from the one displayed in figure \ref{fig:prat} as a result of tidal migration.  However, this is not to say that the orbital period ratios of a given pair of planets cannot evolve substantially.  In our reference model, depending on the initial lay down of semi-major axes, the orbits of the first 2-3 planets typically diverge, while those of our TRAPPIST-1d-f analogs converge with one another, and the furthest 2-3 planets often experience negligible semi-major axis evolution.  Thus, it is important to note that these effects essentially cancel each other out when interpreting the difference between the red and black lines in the top panel of figure \ref{fig:tides}.  Indeed, the inner two planets' orbital period ratios change by 2-5$\%$ in the majority of our simulations, and those of the second and third planets typically evolve to be 1-2$\%$ less than their initial values.

The second and third panels of figure \ref{fig:tides} show how the eccentricities and inclinations of our analogs are much closer to those of the real TRAPPIST-1 planets after our tidal simulations than before.  In particular, while the majority of our post-formation analogs have eccentricities between 0.01-0.1 (similar to those of the solar system's planets), those values are rapidly reshaped in our \textit{Posidonius} simulations, and the final distribution is in reasonable agreement with that of the real system.  However,  while our analog planets' inclination do marginally de-excite during our follow-on simulations, our model is unable to replicate the remarkable co-planar nature of the actual system of orbits.  However, we note that there is a strong degree of heterogeneity in our sample of systems with respect to their inclination distributions.  Indeed, while only around 40$\%$ of all our final planets have inclinations less that 1$^{\circ}$, 22$\%$ of all our final (post-tidal simulation) systems possess seven planets that all have $i<$ 1.0$^{\circ}$.  In our best simulation, all seven planets have inclinations less than 0.42$^{circ}$, in good agreement with the actual system of planets around TRAPPIST-1.

\begin{figure}
    \centering
    \includegraphics[width=0.49\textwidth]{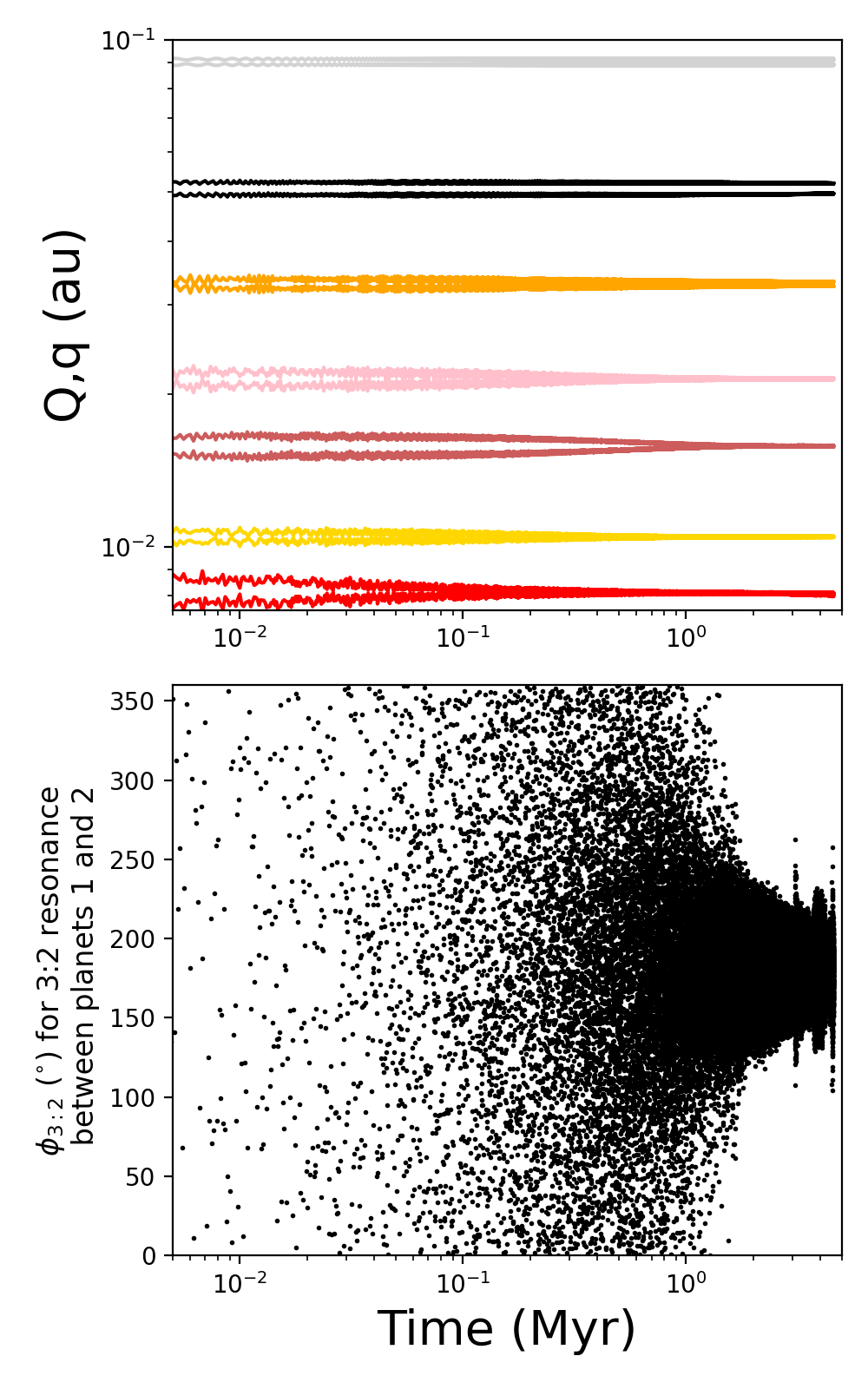}
    \caption{\textbf{Capture of a TRAPPIST-1 analog in the 3:2 MMR after the completion of planet formation.} Top Panel: Example eccentricity evolution in our tidal evolutionary modeling of a TRAPPIST-1 analog system.  The periastron and apastron of each planet is plotted in different colors. Bottom Panel: Resonant Angle for the 3:2 mean motion resonance between the first and second planets in the system.}
    \label{fig:qaq_res_tides}
\end{figure}

The top panel of figure \ref{fig:qaq_res_tides} depicts an example evolution where all seven planets' orbits damp substantially, the final inclination or each planet is less than 1.0$^{\circ}$, and the inner two planets finish the simulation locked in the 3:2 MMR.  In general, we find that resonant capture is a common outcome in our tidal simulations, with just under two resonances being produced in each system (many more planet pairs are briefly trapped in various resonances before being lost).  We utilized the methodology of \citet{clement21_p9} to search for librating resonant angles of the form:
\begin{equation}
	\phi = p \lambda_{2} - q \lambda_{1} - r \varpi_{2} - s \varpi_{1}
	\label{eqn:res}
\end{equation}
\begin{equation}
	p=q+r+s
\end{equation}
where $\varpi$ is the longitude of perihelion and $\lambda$ is the mean longitude of each successive planet.  Of the resonances we detected that are stable at the end of the simulation, 14\% are in 3:2, 82\% inhabit the 2:1, and the remaining 4$\%$ are higher order.  The majority of our higher order captures are in the 5:3 and 8:5 MMRs, however we did find a single instance of capture in the 7:4.

While our follow-on simulations did not produce any resonant chains resembling TRAPPIST-1's, nor did they yield any final configurations where all seven planets were resonant, our results are encouraging in the sense that they demonstrate several instances of delayed capture in higher order resonances that are similar to those between the inner three planets in the real system.  This suggests all six MMRs in the system might not necessarily be primordial.  If this were the case, one could envision a scenario where an initial system of $\geq$ 7 planets form in a chain of 3:2 resonances, before experiencing a delayed dynamical instability \citep{izidoro17,izidoro19} that dislodges the inner planets from resonances and potentially consolidates worlds and creates the distinctively large mass ratio between planets c and d \citep[see][for a similar version of this model applied to the TRAPPIST-1 system]{childs23}.  Our work demonstrates how subsequent tidal migration in the aftermath of such a series of events might be enough to reconstitute a seven planet resonant chain that is not entirely comprised of the 3:2 or 2:1 MMRs.  This scenario is somewhat similar to the disk edge recession framework of \citet{pichierri24} in the sense that it invokes resonant chain formation in two phases.

\section{Conclusions}

In this paper, we analyzed a large collection of N-body simulations of the formation of the TRAPPIST-1 exoplanets.  Our models use an integrator that allows for collisional fragmentation \citep{chambers13}.  We selected this methodology in order to assess the likelihood that the apparent negative gradient of uncompressed densities in the real system is reflective of a radial trend in core mass fraction (CMF), volatile content, or both.  Given the high degree of radial mixing that occurs during the planet formation process, even for the most extreme presumed initial volatile lay downs, we found it challenging to produce a strong final radial trend in bulk planetary water mass fractions (WMF).  Simply put, our models predict that all seven planets are likely to have very similar WMFs.  Radial trends in planet CMF emerged more readily in our models; however, they tended to be more extreme than what is needed to explain the hierarchical distribution of surface gravities in the real system \citep{agol21}.  Since this discrepancy could simply be the result of mantle and core material exchange being less efficient in a real planet-forming disk than in our models, the major conclusion of our study is that the TRAPPIST-1 planets likely have similar volatile inventories, and varied bulk Iron-Silicate ratios.

The primary initial parameter varied in our planet formation simulations was the steepness of the disk's radial surface density profile (the surface density in our disk's are proportional to $r^{\alpha}$, and we tested values of $\alpha=$ -1.0, -1.5 and -2.5).  While none of our simulations were capable of consistently matching the precise ``bi-modal'' mass configuration of the real system, our $\alpha=$ -2.5 batch of simulations was most successful as it more frequently produced systems of 6-8 total planets consisting of two $\gtrsim$ 1.0 $M_{\oplus}$ interior worlds, slightly smaller versions of planets e-g ($\lesssim$ 1.0 $M_{\oplus}$) and a small TRAPPIST-1h.  Unlike our disks with shallower surface density profiles, these $\alpha=$ -2.5 disk's did not over-populate the region exterior to TRAPPIST-1h with debris in a manner that would be inconsistent with ALMA-derived upper limits on the total mass of an undetected hot debris disk in the system \citep{marino20}.  Finally, our $\alpha=$ -2.5 models were more successful than our other simulations in terms of their ability to produce hierarchical distributions of planet CMFs and WMFs.

We concluded our study by evaluating follow-on simulations of our most successful systems that used a code that accounts for tidal effects and rotational flattening \citep{blanco-cuaresma17,bolmont20,blanco--cuaresma21}.  Through this process, we produced many chains of multiple planets in mean motion resonances (MMR) with one another.  While the vast majority of the systems we tested became entrapped in the dominant first order resonances (3:2 and 2:1), we demonstrated several captures in higher order resonances like the ones present in the real TRAPPIST-1 system (8:5 and 5:3).  However, none of our formed resonant chains perfectly resembled the observed one.

In this paper we used a specific methodology and a targeted set of assumptions to investigate whether collisional fragmentation and long-term tidal evolution might be responsible for the apparent gradient of surface gravities and higher order resonances observed in the real TRAPPIST-1 system.  Future work should continue to build on our results, and those of other works in the literature that account for gas dynamics and pebble accretion \citep[neglected here for, e.g.][]{ormel17,coleman19} when interrogating the formation of this fascinating system.

\section*{Acknowledgments}

The authors thank an anonymous reviewer for an insightful and constructive review that significantly improved the quality of this manuscript.  We also thank Tim Lichtenberg, John Chambers and Billy Quarles for fruitful discussions and insight that greatly improved the manuscript and the presentation of the results. This material is based upon work performed as part of the CHAMPs (Consortium on Habitability and Atmospheres of M-dwarf Planets) team, supported by the National Aeronautics and Space Administration (NASA) under Grant Nos. 80NSSC21K0905 and 80NSSC23K1399 issued through the Interdisciplinary Consortia for Astrobiology Research (ICAR) program.  MSC is supported by NASA Emerging Worlds grant 80NSSC23K0868 and NASA’s CHAMPs team.  The computations presented here were supported by the Carnegie Institution and were conducted in the Resnick High Performance Computing Center, a facility supported by Resnick Sustainability Institute at the California Institute of Technology.

\bibliographystyle{apj}
\bibliography{mdwarf.bib}
\end{document}